\documentclass[10pt, conference]{IEEEtran}
\IEEEoverridecommandlockouts
\usepackage{kotex}
\usepackage{ifpdf}
\usepackage{cite}
\usepackage{bbm}
\ifCLASSINFOpdf
\usepackage{graphicx}
\usepackage{subfigure}
\usepackage{epstopdf}
\usepackage{xcolor}
\else
\fi
\usepackage{float}
\usepackage{amsmath}
\usepackage{amssymb}
\usepackage{amsthm}
\usepackage{mathtools}

\newtheorem{lemma}{Lemma}

\newtheorem{thm}{Theorem}

\newtheorem{assume}{Assumption}
\usepackage{algorithm}
\usepackage{algpseudocode}

\allowdisplaybreaks
\usepackage{array}
\usepackage{stfloats}
\usepackage{tikz}
\theoremstyle{definition}

\usepackage[shortlabels]{enumitem}
\usepackage{comment}
\usepackage{multirow}
\usepackage{booktabs}

\def\BibTeX{{\rm B\kern-.05em{\sc i\kern-.025em b}\kern-.08em
    T\kern-.1667em\lower.7ex\hbox{E}\kern-.125emX}}
\begin{document}
\title{Feature Reconstruction Aided Federated Learning for Image Semantic Communication\\
\thanks{This work was supported by Institute of Information \& communications Technology Planning \& Evaluation (IITP) grant funded by the Korea government(MSIT). (No.RS-2024-00398948, Next Generation Semantic Communication Network Research Center)}
}

\author{\IEEEauthorblockN{Yoon Huh\IEEEauthorrefmark{1}\IEEEauthorrefmark{2}, Bumjun Kim\IEEEauthorrefmark{1}\IEEEauthorrefmark{2}, and Wan Choi\IEEEauthorrefmark{1}\IEEEauthorrefmark{2}}
\IEEEauthorblockA{\IEEEauthorrefmark{1}Department of Electrical and Computer Engineering, Seoul National University, Seoul 08826, South Korea}
\IEEEauthorblockA{\IEEEauthorrefmark{2}Institute of New Media and Communications, Seoul National University, Seoul 08826, South Korea}
\{mnihy621, eithank96, wanchoi\}@snu.ac.kr
}

\maketitle

\begin{abstract}
Research in semantic communication has garnered considerable attention, particularly in the area of image transmission, where joint source-channel coding (JSCC)-based neural network (NN) modules are frequently employed. However, these systems often experience performance degradation over time due to an outdated knowledge base, highlighting the need for periodic updates.  To address this challenge in the context of training JSCC modules for image transmission, we propose a federated learning (FL) algorithm with semantic feature reconstruction (FR), named FedSFR. This algorithm more efficiently utilizes the available communication capacity by allowing some of the selected FL participants to transmit smaller feature vectors instead of local update information. Unlike conventional FL methods, our approach integrates FR at the parameter server (PS), stabilizing training and enhancing image transmission quality. Experimental results demonstrate that the proposed scheme significantly enhances both the stability and effectiveness of the FL process compared to other algorithms. Furthermore, we mathematically derive the convergence rate to validate the improved performance.
\end{abstract}


\section{Introduction}
To enable more spectral efficient communications in sixth-generation (6G) communication, which is evolving into task-oriented communication \cite{shi2023task}, semantic communication \cite{luo2022semantic} is gaining significant attention. This approach efficiently transmits only task-relevant data and effectively manages various types of source data. One notable application is the image reconstruction task, which is the focus of this paper and is referred to as image semantic communication (ISC). 

Semantic communication often relies on joint source-channel coding (JSCC) techniques \cite{huh2025universal}.  Typically, JSCC is implemented using neural networks (NNs) and employs an autoencoder structure, consisting of an encoder and a decoder, in ISC. It optimizes source and channel coding together, reducing redundancy in the transmitted data and improving overall efficiency. However, over time, the underlying distribution of the knowledge base \cite{luo2022semantic}, represented by the training dataset, gradually shifts\footnote{While continual learning techniques help mitigate distribution shifts like catastrophic forgetting, our focus in this paper is on developing an FL framework rather than addressing continual learning issues.}. This shift leads to a steady decline in performance, especially in dynamic environments where data distributions change. As a result, continuous updates to the model are essential to maintain optimal performance.

Federated learning (FL) has emerged as a vital solution to these challenges \cite{sun2024federated, xu2024federated}, which facilitates the periodic updating of NN-based JSCC models by aggregating knowledge from multiple distributed sources while preserving data privacy. In this framework, ISC users act as FL clients, each equipped with both a JSCC encoder and decoder, which serve as the transmitter and receiver, respectively, during ISC and together form the local model during FL. The parameter server (PS) aggregates the global JSCC model as in conventional FL, however, ISC occurs exclusively among the clients, excluding the PS, after FL.

Nevertheless, FL also faces significant challenges \cite{kairouz2021advances}, particularly in environments with limited communication resources. The process of aggregating updates from multiple clients can lead to substantial communication overhead since they transmit the whole local model to the PS. To alleviate these concerns, sparsification techniques \cite{stich2018sparsified} have been employed, where only the most significant values in local models are transmitted. Additionally, error feedback strategies \cite{karimireddy2019error}, i.e., storing the error from sparsification and incorporating them into future model updates, are used to compensate for the information loss due to sparsification to improve the overall efficiency of the FL process in capacity-constrained networks.

Building on this, several studies have explored FL frameworks specifically designed for semantic communication. For example, the authors of \cite{sun2024federated} proposed an FL framework that optimizes JSCC model for image classification tasks. Some clients send their local models,  while the others transmit output feature vectors from the encoder, along with the corresponding class information, to the PS. The PS updates the previous global model through knowledge distillation using the received feature vectors and ground-truth labels, and then aggregates the updated model with the received local models. However, since the class label in classification tasks corresponds to the original image for image reconstruction tasks, this approach incurs a privacy risk and is thus not applicable in ISC. In contrast, \cite{xu2024federated} extended the FL framework to handle image reconstruction. Instead of transmitting the local models, clients exchange output feature vectors from the encoder and intermediate feature vectors from the decoder with the PS. The model components corresponding to these feature vectors are updated via distillation at both the clients and the PS, while the remaining parts are optimized locally at the clients. Since the PS relies solely on feature vectors without direct local model exchanges via FL, forming a common global model becomes challenging. As a result, the encoder produces inconsistent feature vectors across the clients, inherently limiting performance without leveraging the benefits of local model averaging.

To address these challenges, we propose an advanced FL algorithm with semantic feature reconstruction (FedSFR). Our key contributions are as follows:
\begin{itemize}
    \item We introduce FedSFR, which divides clients into two groups: one sends local model updates as in conventional FL, while the other transmits compressed feature vectors from the JSCC encoder to the PS.  After the model aggregation, the PS further refines the model through additional feature reconstruction (FR) learning, where the received feature vectors are sequentially processed through the decoder and then the encoder—differing from client-side processing. This approach combines the benefit of autoencoder-based ISC, by reversing the model’s data processing order, with that of local model averaging in FL, leading to enhanced feature representation and improved model consistency.
    
    \item FedSFR enables clients to adaptively select between transmitting conventional local updates or compact feature vectors based on their channel quality. This adaptive mechanism significantly reduces communication overhead while ensuring stable and efficient training.
       
    \item We provide a comprehensive analysis of the proposed method, including its convergence bound, convergence rate, and the ability to mitigate sparsification errors through server-side updates.
    
    \item Extensive experiments on both low-resolution and high-resolution datasets demonstrate that FedSFR substantially improves training stability and convergence speed compared to baseline methods, validating both its effectiveness and our theoretical analysis.
\end{itemize}

\section{System Model}\label{sec:system model}
This section presents the foundational framework for semantic communication and federated learning, which are central concepts explored in this paper.  After optimizing the NN parameters for the JSCC encoder and decoder at the client through the wireless FL process, ISC is carried out between clients. In this setup, semantic communication users participates in FL as clients, working collaboratively with the PS. The commonly used notations are summarized in TABLE \ref{tab:Notations about FL at global iteration t}. 

\begin{table}[t!]
\caption{Notations about FL at global iteration $t$.}
\label{tab:Notations about FL at global iteration t}
\centering
\begin{tabular}{cl}
\hline
\textbf{Notation}                                    & \textbf{Description}                                                                                     \\ \hline
$\boldsymbol{w}^{\scriptstyle(t)}$                   & Global model to be downloaded at the clients                                                             \\
$\boldsymbol{w}^{\scriptstyle(t + \frac{1}{2})}$     & Global model to be updated at the PS                                                                     \\
$\boldsymbol{w}_k^{\scriptstyle(t,e)}$               & Local model of client $k$ at local iteration $e$                                                         \\
$\boldsymbol{w}_s^{\scriptstyle(t + \frac{1}{2},e)}$ & Server model at server iteration $e$                                                                     \\
$\mathbf{g}_k^{\scriptstyle(t)}$                     & Local update information at client $k$                                                                   \\
$\mathbf{m}_k^{\scriptstyle(t)}$                     & Error memory before sparsification at client $k$                                                         \\
$\mathcal{P}_k$                                      & Shared public dataset at client $k$                                                                      \\
$\mathcal{Y}_k^{\scriptstyle(t)}$                    & Set of feature vectors after local update at client $k$                                                  \\
$N$                                                  & Number of JSCC model parameters                                                                          \\
$K$ ($\mathcal{A}$)                                  & Number (Set) of total clients                                                                            \\
$K_m$ ($\mathcal{A}_m^{\scriptstyle(t)}$)            & Number (Set) of clients sending local update information                                                 \\
$K_o$ ($\mathcal{A}_o^{\scriptstyle(t)}$)            & Number (Set) of clients sending feature vectors                                                          \\
$S_m$ / $S_o$                                        & Sparsification level of clients in $\mathcal{A}_m^{\scriptstyle(t)}$ / $\mathcal{A}_o^{\scriptstyle(t)}$ \\
$T$                                                  & Number of global iterations                                                                              \\
$E_c$                                                & Number of local iterations                                                                               \\
$E_s$                                                & Number of server iterations                                                                              \\ \hline
\end{tabular}
\vspace{-4mm}
\end{table}

\subsection{Image Semantic Communication}
Consider a JSCC-based semantic communication system for image transmission, where the objective is image reconstruction. The communication process proceeds as follows. The transmitter encodes a source image $\mathbf{X} \in \mathbb{R}^{C \times H \times W}$, where $C$, $H$, and $W$ represent the number of channels, height, and width of the image, respectively, into a feature vector $\mathbf{y} \in \mathbb{R}^d$ using an encoder $f_{\boldsymbol{\theta}}$, which is parameterized by $\boldsymbol{\theta}$. This encoding can be expressed as $\mathbf{y} = f_{\boldsymbol{\theta}}(\mathbf{X})$.

The feature vector $\mathbf{y}$ is then normalized as $\Tilde{\mathbf{y}} = \mathbf{y}/\|\mathbf{y}\|_2$ and transmitted over an additive white Gaussian noise (AWGN) channel. At the receiver, the noisy feature vector is decoded to reconstruct the image $\hat{\mathbf{X}} \in \mathbb{R}^{C \times H \times W}$ using a decoder $f^{-1}_{\boldsymbol{\phi}}$, parameterized by $\boldsymbol{\phi}$. This process is described as $\hat{\mathbf{X}} = f^{-1}_{\boldsymbol{\phi}}(\Tilde{\mathbf{y}} + \mathbf{n})$, where $\mathbf{n} \sim \mathcal{N}(\boldsymbol{0}_d, \sigma^2 \mathbf{I}_d)$ represents the noise, and the signal-to-noise ratio (SNR) is given by $1/\sigma^2$.

Define the learnable parameter $\boldsymbol{w} = \{\boldsymbol{\theta},\boldsymbol{\phi}\} \in \mathbb{R}^{N}$, where $N$ represents the number of NN parameters in the JSCC model. $\boldsymbol{w}$ is trained using a loss function $F(\boldsymbol{w}) = \frac{1}{|\mathcal{D}|}\sum_{\mathbf{X}\in\mathcal{D}}l_c(\boldsymbol{w}; \mathbf{X})$, which serves as the global objective function at the PS in FL. Here, $\mathcal{D}$ denotes the dataset containing the source images, and the loss function $l_c(\boldsymbol{w}; \mathbf{X}) = \textsf{MSE}(\hat{\mathbf{X}}, \mathbf{X})$ represents the mean squared error (MSE) between the original image $\mathbf{X}$ and the reconstructed image $\hat{\mathbf{X}}$.

\subsection{Federated Learning}
Consider a wireless FL, based on the FedAvg algorithm \cite{mcmahan2017communication}, where the PS optimizes the global model for both the  JSCC encoder and decoder, $f_{\boldsymbol{\theta}}$ and $f^{-1}_{\boldsymbol{\phi}}$, in collaboration with $K$ clients in the set $\mathcal{A}$, i.e., $|\mathcal{A}| = K$. Each client $k$ possesses a local dataset $\mathcal{D}_k$, such that the global dataset is given by $\mathcal{D} = \bigcup_{k \in \mathcal{K}} \mathcal{D}_k$.The local objective function for each client can be expressed as $F_k(\boldsymbol{w}) = \frac{1}{|\mathcal{D}_k|}\sum_{\mathbf{X}\in\mathcal{D}_k}l_c(\boldsymbol{w}; \mathbf{X})$, resulting in the global objective function $F(\boldsymbol{w}) = \sum_{k\in\mathcal{A}}p_k F_k(\boldsymbol{w})$ where $p_k = |\mathcal{D}_k| / |\mathcal{D}|$ represents the proportion of the global dataset owned by client $k$.

FL process primarily consists of two main components: the local update process and the global update process. In the local update process, each client receives the global model $\boldsymbol{w}_k^{\scriptstyle(t, 0)} = \boldsymbol{w}^{\scriptstyle(t)} \in \mathbb{R}^{N}$ from the PS at global iteration $t \in \{0, \dots, T - 1\}$. The client then locally updates this model using a mini-batch-based stochastic gradient descent (SGD) algorithm with the dataset $\mathcal{D}_k$. The update rule is given by
\begin{align}\label{eq:local iteration}
    \boldsymbol{w}_k^{\scriptstyle(t, e + 1)} = \boldsymbol{w}_k^{\scriptstyle(t, e)} - \eta_c^{\scriptstyle(t)} \nabla F_k^{\scriptstyle(t, e)}(\boldsymbol{w}_k^{\scriptstyle(t, e)}), 
\end{align}
where $e \in \{0, \dots, E_c - 1\}$ denotes the local iteration number, $\eta_c^{\scriptstyle(t)}$ represents the local learning rate at global iteration $t$, and $E_c$ is the total number of local iterations. The local gradient vector at iteration $e$ is defined as $\nabla F_k^{\scriptstyle(t, e)}(\boldsymbol{w}_k^{\scriptstyle(t, e)}) = \frac{1}{|\mathcal{D}_k^{\scriptstyle(t, e)}|} \sum_{\mathbf{X} \in \mathcal{D}_k^{\scriptstyle(t, e)}} \nabla l_c(\boldsymbol{w}_k^{\scriptstyle(t, e)}; \mathbf{X})$, 
where $\mathcal{D}_k^{\scriptstyle(t, e)}$ represents a mini-batch sampled from $\mathcal{D}_k$ at local iteration $e$.

Then, the client sends the local update information $\mathbf{g}_k^{\scriptstyle(t)}\in\mathbb{R}^{N}$ to the PS. Suppose that only the clients in $\mathcal{A}_m^{\scriptstyle(t)}$ send $\mathbf{g}_k^{\scriptstyle(t)}$ via the uplink transmission, where $\mathcal{A}_m^{\scriptstyle(t)}\subset\mathcal{A}$ is a subset of the participating clients at global iteration $t$ and $|\mathcal{A}_m^{\scriptstyle(t)}| = K_m$. Note that the subscript `$m$' stands for local \underline{m}odel. Assuming the utilization of top-$S$ sparsification\footnote{Considering the channel capacity between the PS and the client, we must control the amount of transmitted gradient information to ensure reliable communication, i.e., error-free transmission. This means that $S$ may vary across clients depending on their channel conditions. The top-$S$ sparsification algorithm selects the $S$ largest values in the vector by ranking elements based on their magnitudes. Here, we apply sparsification on a per-layer basis in NN.} with error feedback strategy \cite{amiri2020federated}, the local update information is expressed as
\begin{align}\label{eq:sparsification}
    \mathbf{g}_k^{\scriptstyle(t)} = \textsf{Sparse}(\mathbf{m}_k^{\scriptstyle(t)} + \eta_c^{\scriptstyle(t)}\sum_{e = 0}^{E_c - 1}\nabla F_k^{\scriptstyle(t, e)}(\boldsymbol{w}_k^{\scriptstyle(t, e)})),
\end{align}
where $\textsf{Sparse}$ operates the top-$S$ sparsification algorithm and $\mathbf{m}_k^{\scriptstyle(t)}\in\mathbb{R}^{N}$ is the error memory at the client $k$ at global iteration $t$, which is updated as
\begin{align}\label{eq:error memory update}
    \mathbf{m}_k^{\scriptstyle(t + 1)} = \mathbf{m}_k^{\scriptstyle(t)} + \eta_c^{\scriptstyle(t)}\sum_{e = 0}^{E_c - 1}\nabla F_k^{\scriptstyle(t, e)}(\boldsymbol{w}_k^{\scriptstyle(t, e)}) - \mathbf{g}_k^{\scriptstyle(t)}.
\end{align}
$\mathbf{m}_k^{\scriptstyle(t)}$ retains the differences between the full gradient and the transmitted sparse gradient. This mechanism helps gradually correct the loss of information over time, ultimately improving the accuracy and convergence of the global model. The sparsification level $S_k$ can vary among clients and is selected based on the channel capacity of each client.

Note that the PS views the entire NN model as a composite function, i.e., $f_{\boldsymbol{\theta}}\circ f^{-1}_{\boldsymbol{\phi}}$, where the JSCC encoder is applied after the JSCC decoder in sequence. Moreover, the model at the PS is regarded as a unified entity, rather than two separate models (the decoder and encoder) connected. In contrast, the clients can differentiate between the encoder and decoder components from the global model in a format of the PS, and send their local update information in the same format to the PS.

During the global update process, assume that the PS receives the local update information $\{\mathbf{g}_k^{\scriptstyle(t)}\}_{k\in\mathcal{A}_m^{\scriptstyle(t)}}$ without uplink transmission error since the sparsification level is selected based on the channel capacity of each client. Then, the PS updates the global model by weighted averaging \cite{li2019convergence}:
\begin{align}\label{eq:global update}
    \boldsymbol{w}^{\scriptstyle(t + 1)} = \boldsymbol{w}^{\scriptstyle(t)} - \frac{K}{K_m}\sum_{k\in\mathcal{A}_m^{\scriptstyle(t)}}p_k \mathbf{g}_k^{\scriptstyle(t)}.
\end{align}
After the aggregation, the PS broadcasts the global model $\boldsymbol{w}^{\scriptstyle(t + 1)}$ to the clients via the downlink transmission.

\section{FL with Semantic Feature Reconstruction}\label{sec:fl with semantic feature reconstruction}
To efficiently utilize the limited communication resources, we consider that the model size $N$ is typically much larger than the feature vector size $d$ in ISC, i.e., $N \gg d$. Our strategy leverages the feature vectors from clients with poor channel conditions, which brings the effect of \textit{transmitting the local update information to the PS}.  Given the reconstruction tasks, feature vectors serve as a compressed yet highly informative summary of local data and models, capturing essential patterns and characteristics. This approach is analogous to federated knowledge distillation in classification tasks, where logit vectors from a client encapsulate information about the local classifier \cite{seo202216}. Instead of directly transmitting local gradient updates, our strategy utilizes feature vectors extracted from the most recent updates of the local JSCC encoder.

\begin{figure}[!t]
    \centering
    \includegraphics[width=0.9\linewidth]{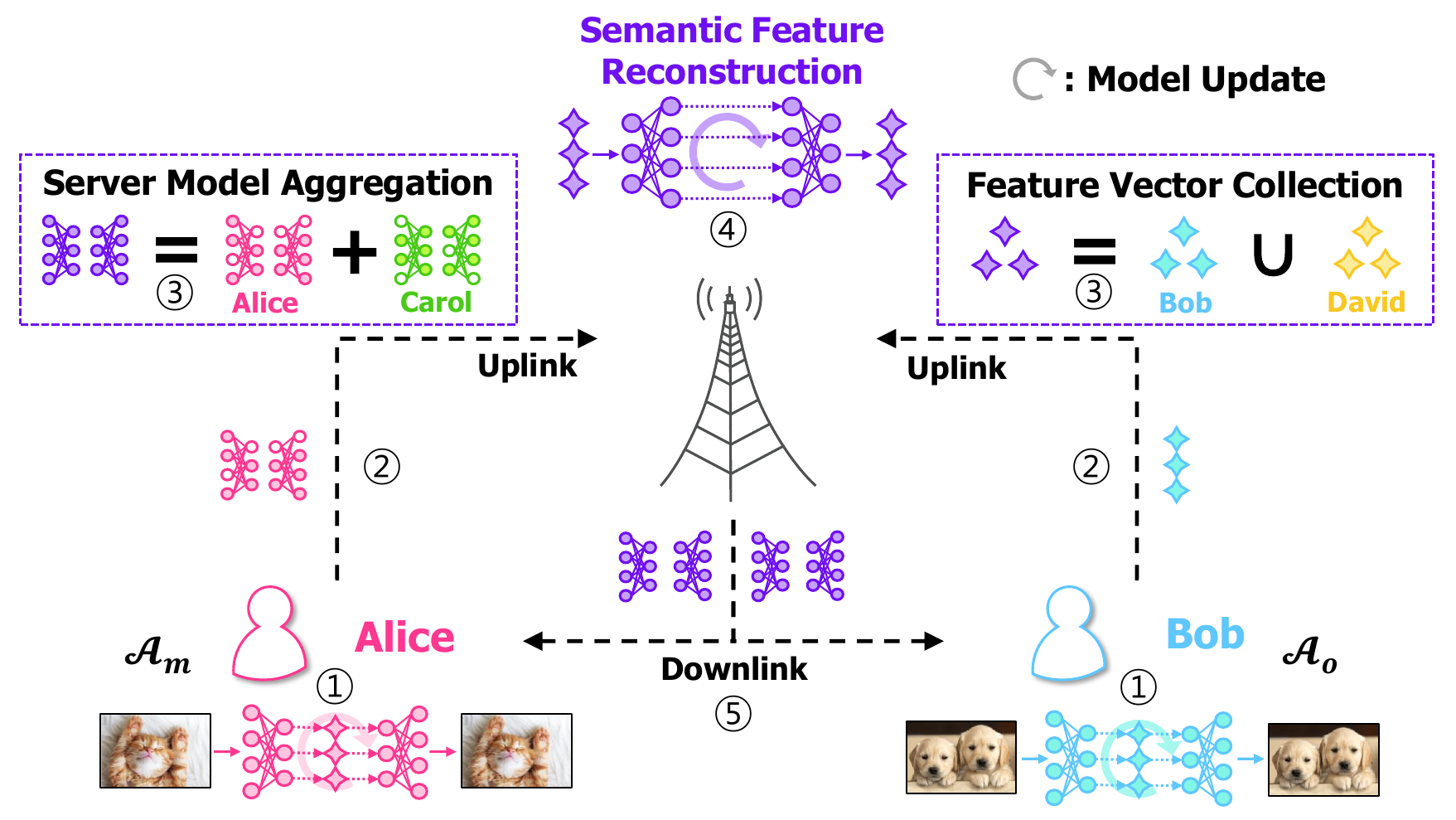}
    \vspace{-3mm}
    \caption{Overall procedure of FedSFR with the numbered algorithmic steps.}
    \label{fig:Overall procedure of FedSFR with the numbered algorithmic steps}
    \vspace{-5mm}
\end{figure}

As illustrated in Fig. \ref{fig:Overall procedure of FedSFR with the numbered algorithmic steps}, which outlines the five numbered \texttt{steps}, the overall procedure of the FedSFR algorithm is as follows. Consider two sets of participating clients, denoted as $\mathcal{A}_m^{\scriptstyle(t)}$ and $\mathcal{A}_o^{\scriptstyle(t)}$, where clients in $\mathcal{A}_m^{\scriptstyle(t)}$ have better channel conditions than those in $\mathcal{A}_o^{\scriptstyle(t)}$. Note that the subscript `$o$' represents the encoder \underline{o}utput feature vector.  After the local update process given in \eqref{eq:local iteration} (\texttt{step 1}), each client $k$ in $\mathcal{A}_m^{\scriptstyle(t)}$ transmits the local update information $\mathbf{g}_k^{\scriptstyle(t)}$ about the local model according to  \eqref{eq:sparsification} and \eqref{eq:error memory update} (\texttt{step 2}), with $S_k = S_m$. Meanwhile, each client $k$ in $\mathcal{A}_o^{\scriptstyle(t)}$, where $|\mathcal{A}_o^{\scriptstyle(t)}| = K_o$, transmits a set $\mathcal{Y}_k^{\scriptstyle(t)}$ containing multiple instances of $\mathbf{y}$ (\texttt{step 2}). Here, $\mathbf{y}$ is generated from the JSCC encoder $f_{\boldsymbol{\theta}}$ with $\boldsymbol{\theta} = \boldsymbol{\theta}_k^{\scriptstyle(t, E_c)}$, and $S_k$ is set to $S_o$. Although $S_o$ does not represent sparsification but rather the amount of feature vectors, we abuse the notation $S$ to indicate the transmit data size.

Assume that $\mathcal{Y}_k^{\scriptstyle(t)}$ is derived from a shared public dataset $\mathcal{P}_k$ \cite{zhao2018federated}, which varies across clients and is also used as part of the local dataset, i.e., $\mathcal{P}_k \subset \mathcal{D}_k$.  By randomly selecting only a subset of $\mathcal{P}_k$, the total data size of $\mathcal{Y}_k^{\scriptstyle(t)}$ remains less than or equal to $S_o$, where $S_o < S_m$. Next, the PS aggregates $\{\mathbf{g}_k^{\scriptstyle(t)}\}_{k\in\mathcal{A}_o^{\scriptstyle(t)}}$ and construct $\mathcal{D}_s^{\scriptstyle(t + \frac{1}{2})} = \bigcup_{k \in \mathcal{A}_o^{\scriptstyle(t)}} \mathcal{Y}_k^{\scriptstyle(t)}$ by collecting feature vectors (\texttt{step 3}). In contrast to \eqref{eq:global update}, the weighted averaged global model is expressed as
\begin{align}\label{eq:global aggregation}
    \boldsymbol{w}^{\scriptstyle(t + \frac{1}{2})} = \boldsymbol{w}^{\scriptstyle(t)} - \frac{K}{K_m}\sum_{k\in\mathcal{A}_m^{\scriptstyle(t)}}p_k \mathbf{g}_k^{\scriptstyle(t)}.
\end{align}

The PS then further updates the global model, $\boldsymbol{w}_s^{\scriptstyle(t + \frac{1}{2}, 0)} = \boldsymbol{w}^{\scriptstyle(t + \frac{1}{2})}$, by learning to reconstruct the feature vectors from clients in $\mathcal{A}_o^{\scriptstyle(t)}$ (\texttt{step 4}). Since the PS possesses the averaged global model—encompassing both the JSCC encoder and decoder—it can directly perform the FR learning using $\mathcal{D}_s^{\scriptstyle(t + \frac{1}{2})}$. This approach differs from the image reconstruction task carried out at the clients during ISC. As a result, the PS effectively \textit{transfers the knowledge of the local model} from each client in $\mathcal{A}_o^{\scriptstyle(t)}$ without requiring the direct transmission of local update information. Given that the PS typically has significantly higher computational capacity than the clients, the additional processing time required for model updates related to the FR task is negligible within the broader FL global iteration process \cite{kairouz2021advances}. If the local update information from client $k \in \mathcal{A}_o^{\scriptstyle(t)}$ is incorporated into the PS through server updates, the error memory is reset as $\mathbf{m}_k^{\scriptstyle(t + 1)} = \boldsymbol{0}_N$.

A data sample $\mathbf{y}$ in $\mathcal{D}_s^{\scriptstyle(t + \frac{1}{2})}$ is reconstructed into $\hat{\mathbf{y}}\in\mathbb{R}^{d}$ via the JSCC decoder and encoder as $\hat{\mathbf{y}} = f_{\boldsymbol{\theta}}(f^{-1}_{\boldsymbol{\phi}}(\Tilde{\mathbf{y}} + \mathbf{n}))$, where $\boldsymbol{\theta} = \boldsymbol{\theta}_s^{\scriptstyle(t + \frac{1}{2}, e)}$, $\boldsymbol{\phi} = \boldsymbol{\phi}_s^{\scriptstyle(t + \frac{1}{2}, e)}$, $\Tilde{\mathbf{y}} = \mathbf{y}/||\mathbf{y}||_2$ and $\mathbf{n} \sim \mathcal{N}(\boldsymbol{0}_d, \sigma^2\mathbf{I}_d)$. Using the same SGD algorithm at the clients, the server iteration process can be expressed as
\begin{align}\label{eq:server iteration}
    \boldsymbol{w}_s^{\scriptstyle(t + \frac{1}{2}, e + 1)} = \boldsymbol{w}_s^{\scriptstyle(t + \frac{1}{2}, e)} - \eta_s^{\scriptstyle(t)}\nabla F_s^{\scriptstyle(t + \frac{1}{2}, e)}(\boldsymbol{w}_s^{\scriptstyle(t + \frac{1}{2}, e)}),
\end{align}
where $e\in\{0, \dots, E_s - 1\}$ is the server iteration number, $E_s$ is the total number of server iterations, and $\eta_s^{\scriptstyle(t)}$ is the server learning rate at global iteration $t$. The server gradient vector at server iteration $e$, utilizing a mini-batch $\mathcal{D}_s^{\scriptstyle(t + \frac{1}{2}, e)}$ sampled from $\mathcal{D}_s^{\scriptstyle(t + \frac{1}{2})}$, is defined as
\begin{align}
\begin{split}
    &\nabla F_s^{\scriptstyle(t + \frac{1}{2}, e)}(\boldsymbol{w}_s^{\scriptstyle(t 
    + \frac{1}{2}, e)})\\
    &\quad\quad = \frac{1}{|\mathcal{D}_s^{\scriptstyle(t + \frac{1}{2}, e)}|}\sum_{\mathbf{y}\in\mathcal{D}_s^{\scriptstyle(t + \frac{1}{2} , e)}}\nabla l_s(\boldsymbol{w}_s^{\scriptstyle(t + \frac{1}{2}, e)}; \mathbf{y}),
\end{split}
\end{align}
where $l_s(\boldsymbol{w}; \mathbf{y}) = \textsf{MSE}(\hat{\mathbf{y}}, \mathbf{y})$. Finally, after the server update process, the PS broadcasts the global model, $\boldsymbol{w}^{\scriptstyle(t + 1)} = \boldsymbol{w}_s^{\scriptstyle(t + \frac{1}{2}, E_s)}$, to the clients (\texttt{step 5}).

\section{Convergence Analysis}\label{sec:convergence analysis}
In this section, we analyze the convergence properties of our proposed scheme. First, we delve into the effect of compensating for the sparsification error, i.e., $\mathbf{m}_k^{\scriptstyle(t + 1)}$, of our proposed scheme. At global iteration $t$, the best global model that the PS can generate is constructed by averaging all the participating clients in $\mathcal{A}^{\scriptstyle(t)} = \mathcal{A}_m^{\scriptstyle(t)}\cup\mathcal{A}_o^{\scriptstyle(t)}$ without sparsification, such as
\begin{align}\label{eq:best iteration}
\begin{split}
    &\boldsymbol{w}^{\scriptstyle(t + 1)} = \boldsymbol{w}^{\scriptstyle(t)} - \frac{K}{K_m + K_o}\\
    &\;\times\sum\nolimits_{k\in\mathcal{A}^{\scriptstyle(t)}}p_k \left(\eta_c^{\scriptstyle(t)}\sum_{e = 0}^{E_c - 1}\nabla F_k^{\scriptstyle(t, e)}(\boldsymbol{w}_k^{\scriptstyle(t, e)}) + \mathbf{m}_k^{\scriptstyle(t)}\right).
\end{split}
\end{align}
On the other hand, in the proposed FL algorithm, the clients in $\mathcal{A}_m^{\scriptstyle(t)}$ send $\mathbf{g}_k^{\scriptstyle(t)}$ and the clients in $\mathcal{A}_o^{\scriptstyle(t)}$ send $\mathcal{Y}_k^{\scriptstyle(t)}$ to the PS such as
\begin{align}\label{eq:proposed iteration}
    &\boldsymbol{w}^{\scriptstyle(t + 1)} = \boldsymbol{w}^{\scriptstyle(t)} - \eta_s^{\scriptstyle(t)}\sum_{e = 0}^{E_s - 1}\nabla F_s^{\scriptstyle(t + \frac{1}{2}, e)}(\boldsymbol{w}_s^{\scriptstyle(t + \frac{1}{2}, e)}) - \frac{K}{K_m}\\
    &\nonumber\;\times\sum_{k\in\mathcal{A}_m^{\scriptstyle(t)}}p_k \left(\eta_c^{\scriptstyle(t)}\sum_{e = 0}^{E_c - 1}\nabla F_k^{\scriptstyle(t, e)}(\boldsymbol{w}_k^{\scriptstyle(t, e)}) + \mathbf{m}_k^{\scriptstyle(t)} - \mathbf{m}_k^{\scriptstyle(t + 1)}\right).
\end{align}

Due to the space constraints, we present the following lemmas and theorem, omitting their proofs.

\begin{lemma}[\!\!\cite{li2019convergence}] \label{lm:client sampling unbiasedness}
If a subset $\mathcal{B}_0$ is uniformly sampled from a given set $\mathcal{B}$ without replacement,
\begin{align*}
    \mathbb{E}_{\mathcal{B}_0}\left[\frac{|\mathcal{B}|}{|\mathcal{B}_0|}\sum_{k\in\mathcal{B}_0}p_k x_k\right] = \sum_{k\in\mathcal{B}}p_k x_k,
\end{align*}
where $\sum_{k\in\mathcal{B}}p_k = 1$.
\end{lemma}

\textbf{Lemma \ref{lm:client sampling unbiasedness}} explains that if the weighted summation of $x_k$ from the subset $\mathcal{B}_0$ is scaled by $|\mathcal{B}|/|\mathcal{B}_0|$, its expectation with respect to the subset sampling is equivalent to the weighted summation of $x_k$ from the total set $\mathcal{B}$. 

Taking the expectation of (\ref{eq:best iteration}) and (\ref{eq:proposed iteration}) with respect to $\mathcal{A}^{\scriptstyle(t)}$, i.e., $\mathbb{E}_{\mathcal{A}^{\scriptstyle(t)}}[\cdot]$, and applying \textbf{Lemma \ref{lm:client sampling unbiasedness}}, the difference between (\ref{eq:best iteration}) and (\ref{eq:proposed iteration}) is $\sum_{k\in\mathcal{A}} p_k \mathbf{m}_k^{\scriptstyle(t + 1)} - \eta_s^{\scriptstyle(t)} \sum_{e = 0}^{E_s - 1} \nabla F_s^{\scriptstyle(t + \frac{1}{2}, e)} (\boldsymbol{w}_s^{\scriptstyle(t + \frac{1}{2}, e)})$. From this difference, our proposed scheme is expected to move closer to the optimal global model described in (\ref{eq:best iteration}) by compensating for the sparsification error through the server iteration process. To quantify the directional similarity between the sparsification error and the server update information, we introduce the following assumption, which includes an inequality.

\begin{assume}\label{as:error memory compensation}
For all $t$, letting $\mathbf{a} = \sum_{k\in\mathcal{A}}p_k\mathbf{m}_k^{\scriptstyle(t + 1)}$ and $\mathbf{b} = \eta_s^{\scriptstyle(t)}\sum_{e = 0}^{E_s - 1}\nabla F_s^{\scriptstyle(t + \frac{1}{2}, e)}(\boldsymbol{w}_s^{\scriptstyle(t + \frac{1}{2}, e)})$, there exists an upper-bound $0 < \epsilon \leq 1$ such that $\frac{\|\mathbf{a} - \mathbf{b}\|_2^2}{\|\mathbf{a}\|_2^2 + \|\mathbf{b}\|_2^2} \leq \epsilon$.
\end{assume}

Since \textbf{Assumption \ref{as:error memory compensation}} implies that $\mathbf{a}^\textsf{T}\mathbf{b} \geq (1 -\epsilon)\|\mathbf{a}\|_2\|\mathbf{b}\|_2$, if $\epsilon$ is sufficiently small, the directions of $\mathbf{a}$ and $\mathbf{b}$ are similar. In other words, since $\mathbf{a}$ represents the sparsification error and $\mathbf{b}$ corresponds to the server update information, the server's iteration process can compensate for the error introduced by the top-$S$ sparsification process.

Next, we derive the convergence bound and rate of our proposed FL algorithm. To this end, we state some assumptions for convergence and a lemma as follows.

\begin{assume}\label{as:smoothness}
The local objective function $F_k(\boldsymbol{w})$ is $\beta_c$-smooth and the global objective function $F(\boldsymbol{w})$ is lower bounded such as $F(\boldsymbol{w}) \geq F(\boldsymbol{w}^*)$ for all $\boldsymbol{w}$.
\end{assume}

\begin{assume}\label{as:local gradient unbiasedness}
The stochastic gradient at each client is unbiased such as $\mathbb{E}[\nabla F_k^{\scriptstyle(t, e)}(\boldsymbol{w})] = \nabla F_k(\boldsymbol{w})$ for all $k, t, e$.
\end{assume}

\begin{assume}\label{as:gradient norm boundedness}
The expected squared norm of the stochastic gradient at each client and the PS is upper bounded such as $\mathbb{E}[\|\nabla F_k^{\scriptstyle(t, e)}(\boldsymbol{w})\|_2^2] \leq G_k^2$ and $\mathbb{E}[\|\nabla F_s^{\scriptstyle(t, e)}(\boldsymbol{w})\|_2^2] \leq G_s^2$.
\end{assume}

\textbf{Assumptions \ref{as:smoothness}, \ref{as:local gradient unbiasedness},  and \ref{as:gradient norm boundedness}} are commonly employed in the mathematical convergence analysis of FL \cite{li2019convergence}. According to \textbf{Assumption \ref{as:smoothness}}, the global objective function $F(\boldsymbol{w})$ is also $\beta_c$-smooth. Throughout this article, assume that all objective functions are non-convex \cite{karimireddy2019error}.

\begin{lemma}\label{lm:error memory norm boundedness}
Under \emph{\textbf{Assumption \ref{as:gradient norm boundedness}}} and \emph{\textbf{Lemma \ref{lm:client sampling unbiasedness}}}, the expected squared norm of the error memory is upper bounded  as
\begin{align*}
    \mathbb{E}[\|\mathbf{m}_k^{\scriptstyle(t + 1)}\|_2^2] \leq \frac{4(1 - \delta)}{\delta^2}(\eta_c^{\scriptstyle(0)})^2 E_c^2 G_k^2
\end{align*}
for all $k$ and $t$, where $0 < \delta = \frac{S}{N} \leq 1$.
\end{lemma}

\begin{thm}\label{thm:convergence rate}
Under \emph{\textbf{Assumptions \ref{as:error memory compensation}, \ref{as:smoothness}, \ref{as:local gradient unbiasedness}, \ref{as:gradient norm boundedness}}}, and \emph{\textbf{Lemmas \ref{lm:client sampling unbiasedness} and \ref{lm:error memory norm boundedness}}}, our proposed FL algorithm with $\eta_c^{\scriptstyle(t)} = \alpha(t)/\sqrt{T}$ and $\eta_s^{\scriptstyle(t)} = \alpha(t)/T^{\frac{3}{4}}$, where $\alpha(t)$ is a monotonic decreasing function with an $\mathcal{O}(1)$ order, has an $\mathcal{O}(1/\sqrt{T})$ order of the convergence rate:
\begin{align*}
\begin{split}
    &\mathbb{E}\left[\frac{1}{T}\sum_{t = 0}^{T - 1}\|\nabla F(\boldsymbol{w}^{\scriptstyle(t)})\|_2^2\right] \leq \frac{1}{\sqrt{T}}\\
    &\quad\times \left(\frac{2}{\alpha(T)}(\mathbb{E}[F(\boldsymbol{w}^{\scriptstyle(0)})] - F(\boldsymbol{w}^*)) + \mathrm{A} + \frac{1}{\sqrt{T}}\mathrm{B} + \frac{1}{T}\mathrm{C}\right),
\end{split}
\end{align*}
where
\begin{align*}
\begin{split}
    \mathrm{A} &= 2\alpha(0)\beta_c\frac{K}{K_m}E_c^2 G_{k, max}^2 + \frac{\alpha(0)^2}{\alpha(T)^2}\frac{E_s^2}{E_c}G_s^2,\\
    \mathrm{B} &= \alpha(0)^2\beta_c^2\left\{\left(\frac{(K - K_m)^2}{K_m^2} + \epsilon\right)\frac{16(1 - \delta)}{\delta^2} + \frac{2}{3}\right\}\\
    &\qquad\times E_c^3 G_{k, max}^2 + 2\frac{\alpha(0)^2}{\alpha(T)}\beta_c E_s^2 G_s^2,\\
    \mathrm{C} &= 4\alpha(0)^2\epsilon\beta_c^2 E_c E_s^2 G_s^2,~and~G_{k, max}^2 = \max_k G_k^2.
\end{split}
\end{align*}
\end{thm}

According to \textbf{Theorem \ref{thm:convergence rate}}, as the right-hand side goes to zero when $T$ becomes large, given the condition $\eta_s^{\scriptstyle(t)} < \eta_c^{\scriptstyle(t)}$, we can conclude that the global model after global iteration $T - 1$, i.e., $\boldsymbol{w}^{\scriptstyle(T)}$, converges in terms of the global objective function $F(\boldsymbol{w})$. This reflects the image reconstruction performance of the JSCC encoder and decoder for ISC. Furthermore, the convergence bound indicates that a smaller parameter $\epsilon$ results in faster convergence. In other words, if the server iteration process effectively compensates for the sparsification error, the optimized global model can be attained more efficiently.

\section{Numerical Results}\label{sec:numerical results}
In this section, we demonstrate the benefits of our proposed FL algorithm for ISC on the CIFAR-10 dataset \cite{krizhevsky2009learning}, which consists of images of size $3\times32\times32$. We evaluate the performance using the PSNR metric. For the simulations, we set $K = 50$, $K_m = 10$, $K_o = 10$, $S_m/N = 0.4$, and $S_o/N = 0.1$ to define the communication environment. The learning rates are initialized as $\eta_s^{\scriptstyle(0)} = 0.001$ and $\eta_c^{\scriptstyle(0)} = 0.01$, with both reduced by a factor of $0.8$ every $10$ global iterations. The mini-batch size is 16 sampled from $\mathcal{D}_k$ with $|\mathcal{D}_k| = 800$, and the number of epochs at the clients and the PS are 3 and 5, respectively. Our JSCC encoder is composed of $5$ convolutional neural network (CNN) layers, and the decoder consists of $5$ transpose CNN layers, together totaling approximately $0.32$M parameters. The output vector size $d$ is 256. Therefore, each client in $\mathcal{A}_o^{\scriptstyle(t)}$ sends $\mathcal{Y}_k^{\scriptstyle(t)}$ with $128 (= |\mathcal{P}_k|)$ feature vectors. The training SNR is fixed at $20$dB, and the PSNR performance is also evaluated at this level.

We consider two baseline methods applicable to FL for ISC: DSGD, which highlights the effectiveness of FR, and FedSFD, which demonstrates the necessity of aggregation. 
\begin{itemize}
\item DSGD \cite{amiri2020federated}: This method uses the top-$S$ sparsification and an error feedback strategy for the participants in $\mathcal{A}^{\scriptstyle(t)}$ without incorporating our proposed server update process based on output feature vectors from some clients. All simulation parameters are identical to those in our FedSFR, except for the parameters related to the PS.
\item FedSFD \cite{xu2024federated}: This method applies only semantic feature distillation, without model aggregation process, for the participants in $\mathcal{A}^{\scriptstyle(t)}$ by utilizing $\{\mathcal{P}_k\}_{k\in\mathcal{A}}$ and transceiving 64 pairs of output feature vectors from the encoder and intermediate feature vectors from the decoder. Due to the need for sufficient training, the uplink data size is larger than that of other methods. The PS hosts a larger JSCC decoder compared to the clients.
\end{itemize}

\begin{figure*}[!t]
    \centering
    \subfigure[CIFAR-10 dataset]{\includegraphics[width=0.31\linewidth]{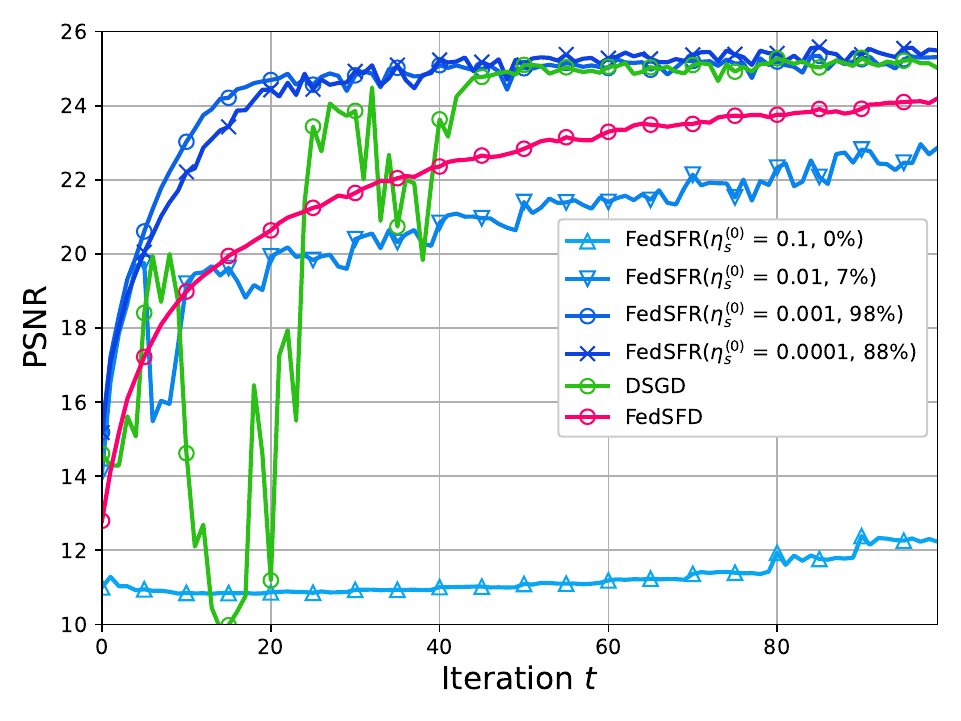}
    \label{fig:PSNR of the proposed scheme and the baselines for CIFAR-10 dataset}}
    \subfigure[Varying $|\mathcal{A}_m^{\scriptstyle(t)}|$ and $|\mathcal{A}_o^{\scriptstyle(t)}|$]{\includegraphics[width=0.31\linewidth]{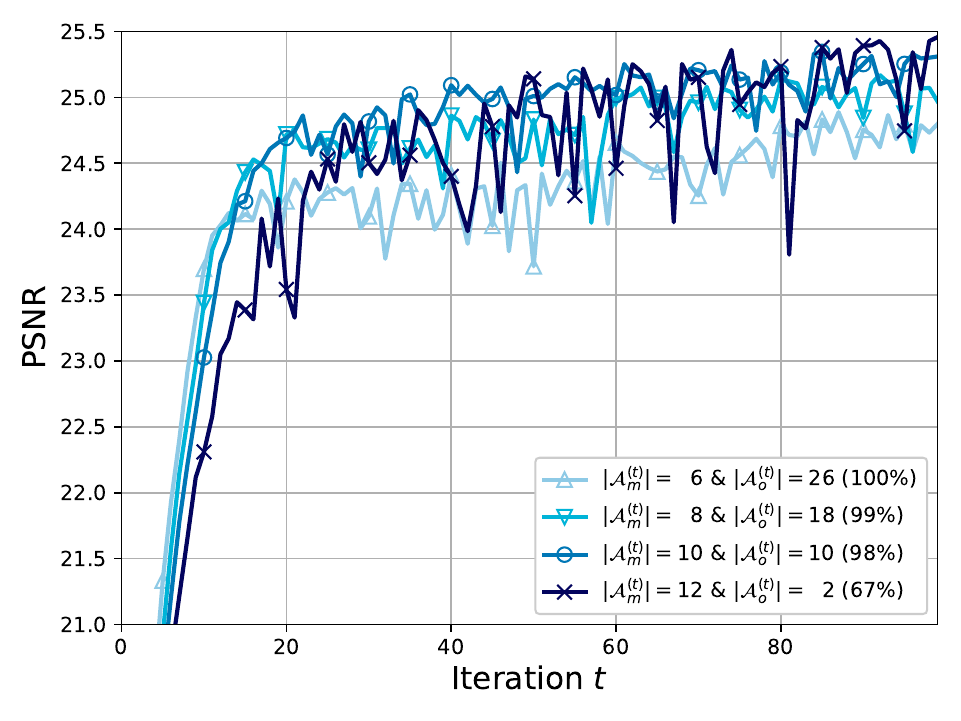}\label{fig:PSNR of the proposed scheme with different server dataset size}}
    \subfigure[CelebA dataset]{\includegraphics[width=0.31\linewidth]{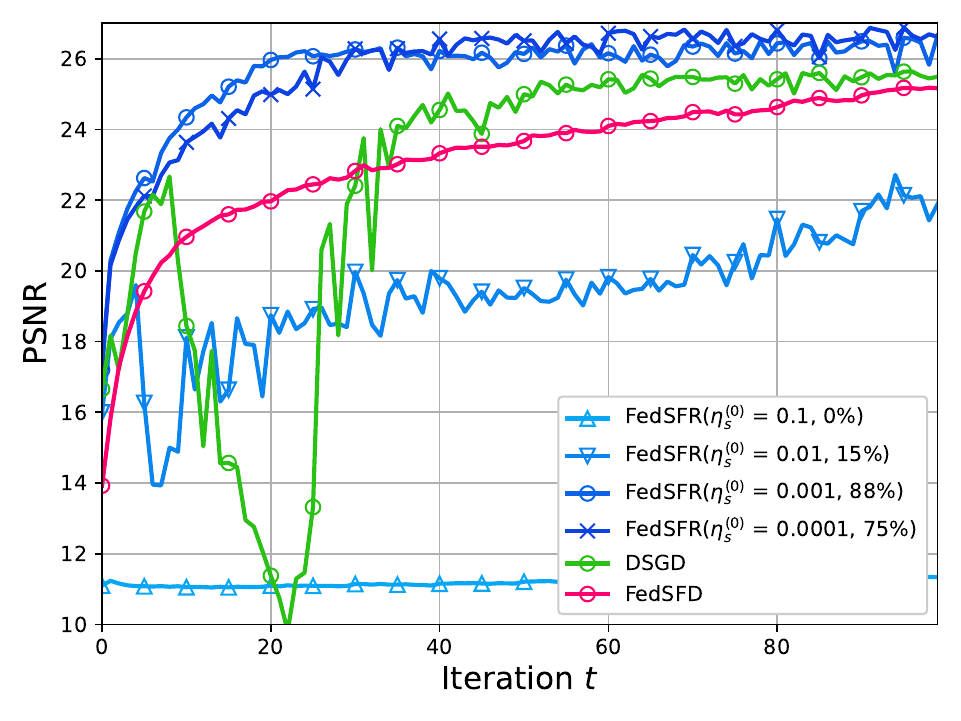}
    \label{fig:PSNR of the proposed scheme and the baselines for CelebA dataset}}
    \vspace{-3mm}
    \caption{PSNR of (a) the proposed scheme and the baselines for CIFAR-10 dataset, (b) the proposed scheme with varying $|\mathcal{A}_m^{\scriptstyle(t)}|$ and $|\mathcal{A}_o^{\scriptstyle(t)}|$ for CIFAR-10 dataset, and (c) the proposed scheme and the baselines for CelebA dataset.}
    \label{fig:Performance comparison}
    \vspace{-5mm}
\end{figure*}

\textbf{Comparison with the Baselines.}\quad   
The three lines with circle markers in Fig. \ref{fig:PSNR of the proposed scheme and the baselines for CIFAR-10 dataset} illustrate the PSNR performance of our proposed scheme compared to the baselines throughout the FL process. DSGD exhibits unstable performance in the early stages of training due to the aggregation of the global model $\boldsymbol{w}^{\scriptstyle(t)}$ using sparsified local updates. This instability arises because, unlike classification tasks—where performance depends on a one-hot vector quantized from softmax-based class probabilities—regression tasks such as image reconstruction require more precise outputs. Additionally, FedSFD demonstrates the poorest performance and the slowest convergence, as its distillation process is not sufficiently effective for FL. In contrast, our proposed scheme, FedSFR, maintains stable training progress, as the PS further refines the aggregated model through FR. This confirms that the server update process effectively mitigates the sparsification error $\mathbf{m}_k^{\scriptstyle(t + 1)}$, as anticipated in \textbf{Assumption \ref{as:error memory compensation}}. This effect is also reflected in the parameter $\epsilon$ within the convergence rate derived in \textbf{Theorem \ref{thm:convergence rate}}. Furthermore, due to its improved stability, the proposed FedSFR achieves significantly faster convergence than the baselines, particularly in the early stages of training.

\textbf{Learning Rate.}\quad The four blue-toned lines in Fig. \ref{fig:PSNR of the proposed scheme and the baselines for CIFAR-10 dataset} illustrates the impact of the initial server learning rate $\eta_s^{\scriptstyle(0)}$, comparing value from the set $\{0.1, 0.01, 0.001, 0.0001\}$. During the FL process, we assess the improvement ratio, shown in the figure’s legend, which represents the proportion of global iterations where FR at the PS results in a performance improvement compared to the model immediately after aggregation, i.e., comparing between $\boldsymbol{w}_s^{\scriptstyle(t + \frac{1}{2}, 0)}$ and $\boldsymbol{w}_s^{\scriptstyle(t + \frac{1}{2}, E_s)}$. Based on \textbf{Theorem \ref{thm:convergence rate}}, we infer that the condition $\eta_s^{\scriptstyle(t)} < \eta_c^{\scriptstyle(t)}$ should be satisfied. Among the tested values, $\eta_s^{\scriptstyle(0)} = 0.1$, which is the worst, and $\eta_s^{\scriptstyle(0)} = 0.01$ yield poor performance, as the overall FL process prioritizes FR over image reconstruction. In contrast, when $\eta_s^{\scriptstyle(0)} = 0.001$ or $\eta_s^{\scriptstyle(0)} = 0.0001$, the training process remains stable. Moreover, these models exhibit a higher improvement ratio and achieve better performance, demonstrating the effectiveness of FR and the necessity of the condition $\eta_s^{\scriptstyle(t)} < \eta_c^{\scriptstyle(t)}$. This confirms that \textbf{Theorem \ref{thm:convergence rate}} effectively guides the selection of the learning rate for FedSFR.

\textbf{$|\mathcal{A}_m^{\scriptstyle(t)}|$ and $|\mathcal{A}_o^{\scriptstyle(t)}|$.}\quad  Fig. \ref{fig:PSNR of the proposed scheme with different server dataset size} exhibits the effect of varying the cardinality of $\mathcal{A}_m^{\scriptstyle(t)}$ and $\mathcal{A}_o^{\scriptstyle(t)}$ while keeping the total communication resources constant. When $|\mathcal{A}_m^{\scriptstyle(t)}| = 6$ and $|\mathcal{A}_o^{\scriptstyle(t)}| = 26$, the performance improves the fastest in the early stage, benefiting from FR with an improvement ratio of 100\%. However, it ultimately converges at the lowest level due to the limited direct aggregation of $\mathbf{g}_k^{\scriptstyle(t)}$. Conversely, when $|\mathcal{A}_m^{\scriptstyle(t)}| = 12$ and $|\mathcal{A}_o^{\scriptstyle(t)}| = 2$, the training process is the slowest at the beginning and less stable due to the small size of $|\mathcal{D}_s|$, leading to an insufficient improvement ratio. Nevertheless, this configuration achieves the highest final performance by directly aggregating more $\mathbf{g}_k^{\scriptstyle(t)}$. Meanwhile, the case of $|\mathcal{A}_m^{\scriptstyle(t)}| = 10$ and $|\mathcal{A}_o^{\scriptstyle(t)}| = 10$ provides the most balanced performance across all aspects.

\textbf{Ablation Studies.}\quad  To further validate our method on higher-resolution images, as shown in Fig. \ref{fig:PSNR of the proposed scheme and the baselines for CelebA dataset}, we conduct additional experiments using the CelebA dataset \cite{liu2015faceattributes}. The results exhibit a similar trend to those observed with the CIFAR-10 dataset, as depicted in Fig. \ref{fig:PSNR of the proposed scheme and the baselines for CIFAR-10 dataset}, highlighting the consistency of FedSFR. Notably, the performance gap between DSGD and our approach increases, indicating that FedSFR is more effective for high-resolution image transmission tasks.
\vspace{-1mm}

\section{Conclusion}\label{sec:conclusion}
In this paper, we introduced FedSFR, which leverages FR at the PS to optimize the NN-based JSCC encoder and decoder for ISC. Expanding upon traditional capacity-limited wireless FL frameworks, our method enhances the global model by minimizing the loss function associated with FR at the PS. This server-side update enables the efficient transfer of ISC capabilities from local models to the global model without transmitting local update information to the PS. Additionally, we derived a convergence rate of $\mathcal{O}(1/\sqrt{T})$ for our FL algorithm, demonstrating how the server update process mitigates the sparsification error. Finally, our simulation results confirmed that the proposed scheme surpasses baseline algorithms in task performance, training stability, and convergence speed.

\bibliographystyle{IEEEtran}  
\bibliography{IEEEabrv,reference}

\end{document}